\documentclass{llncs}
\usepackage{amsfonts}
\usepackage{authblk}
\usepackage[dvips]{graphicx}

\newtheorem{prop}{Proposition}
\newtheorem{lem}{Lemma}
\newtheorem{rem}{Remark}
\title{Negotiating over Bundles and Prices\\Using Aggregate Knowledge\thanks{This is an extended version of our paper with the same title that was accepted for the \emph{$5^{\mathrm{th}}$ International Conference on Electronic Commerce and Web Technologies} (EC-Web 2004), August 30--September 3, 2004 in Zaragoza, Spain In: Bauknecht, Kurt, Martin Bichler \& Birgit Pr\"{o}ll (eds.) \emph{E-Commerce and Web Technologies}, LNCS $3182$, Berlin: Springer, p.\ 218--227.}}
\author{Koye~Somefun\inst{1} \and Tomas~Klos\inst{1} \and Han~La~Poutr\'e\inst{1,2}}
\institute{Center for Mathematics and Computer Science (CWI)\\P.O. Box 94079, 1090 GB Amsterdam, The Netherlands
\and
Eindhoven University of Technology, School of Technology Management,\\P.O. Box 513, 5600 MB Eindhoven, The Netherlands\\
\email{\{koye,tomas,hlp\}@cwi.nl}}

\begin{document}
\maketitle

\begin{abstract}

Combining two or more items and selling them as one good, a practice called bundling, can be a very effective strategy for reducing the costs of producing, marketing, and selling goods. In this paper, we consider a form of multi-issue negotiation where a shop negotiates both the contents and the price of bundles of goods with his customers. We present some key insights about, as well as a technique for, locating mutually beneficial alternatives to the bundle currently under negotiation. The essence of our approach lies in combining historical sales data, condensed into aggregate knowledge, with current data about the ongoing negotiation process, to exploit these insights.

In particular, when negotiating a given bundle of goods with a customer, the shop analyzes the sequence of the customer's offers to determine the progress in the negotiation process. In addition, it uses aggregate knowledge concerning customers' valuations of goods in general. We show how the shop can use these two sources of data to locate promising alternatives to the current bundle. When the current negotiation's progress slows down, the shop may suggest the most promising of those alternatives and, depending on the customer's response, continue negotiating about the alternative bundle, or propose another alternative.

Extensive computer simulation experiments show that our approach increases the speed with which deals are reached, as well as the number and quality of the deals reached, as compared to a benchmark. In addition, we show that the performance of our system is robust to a variety of changes in the negotiation strategies employed by the customers.

\end{abstract}

\section{Introduction}\label{sec:intro}

Combining two or more items and selling them as one good, a practice called bundling, can be a very effective strategy for reducing the costs of producing, marketing, and selling products \cite{Baumol/Etal:1987}. In addition, and maybe more importantly, bundling can stimulate demand for (other) goods or services \cite{Stigler:1963,Schmalensee:1984,Bakos/Brynjolfsson:1999}. To stimulate demand by offering bundles of goods, requires knowledge of customer preferences. Traditionally, firms first acquire such aggregated knowledge about customer preferences, for example through market research or sales data, and then use this knowledge to determine which bundle-price combinations they should offer. Especially for online shops, an appealing alternative approach would be to \emph{negotiate} bundle-price combinations with customers:\footnote{See \cite{kephartFay:2000-acmec,Somefun/Poutre:2003} for other online bundling approaches.} in that case, aggregate knowledge can be used to facilitate an interactive search for the desired bundle and price. Due to the inherently interactive characteristics of negotiation, such an approach can very effectively adapt the configuration of a bundle to the preferences of a customer. A high degree of bundle customization can increase customer satisfaction, which may lead to an increase in the demand for future goods or services.

In this paper, we present an approach that allows a shop to make use of aggregate knowledge about customer preferences. Our procedure uses aggregate knowledge about \emph{many} customers in bilateral negotiations of bundle-price combinations with \emph{individual} customers. Negotiating concerns selecting a subset from a collection of goods or services, viz.\ the bundle, together with a price for that bundle. Thus, the bundle configuration---an array of bits, representing the presence or absence of each of the shop's goods and services in the bundle---together with a price for the bundle, form the negotiation issues. In theory, this is just an instance of multi-issue negotiation. Like the work of~\cite{Klein2003,faratinSierraJennings:2003,Ehtamo2001,somefunGerdingBohtePoutre:2003-amec}, our approach tries to benefit from the so-called win-win opportunities offered by multi-issue negotiation, by finding mutually beneficial alternative bundles during negotiations. The novelty of the approach, however, lies in the use of aggregate knowledge of customer preferences. We show that a bundle with the highest `gains from trade' Pareto-dominates all other bundles within a certain collection of bundles.\footnote{The gains from trade for a bundle are equal to the customer's `valuation' of the bundle minus the seller's valuation of the bundle, which is his (minimum) price. The term refers to the gains obtained from trading the bundle for the price: both sides benefit from trading whenever the customer's valuation is higher than the seller's.}$^,$\footnote{An offer constitutes a Pareto improvement compared to another offer whenever it makes one bargainer better off without making the other worse off. Thus a bundle $b'$ Pareto-dominates another bundle $b$ whenever switching from bundle $b$ to that bundle $b'$ results in a Pareto improvement.}\newcounter{paretoFootnote}\setcounter{paretoFootnote}{\thefootnote} Based on this important insight, we develop an approach for combining aggregate knowledge of customer preferences with data about the ongoing negotiation process, to find alternative bundles that are likely to lead to the highest Pareto improvements. Computer simulations show how, for various types of customers---with distinct negotiation heuristics---our procedure increases the speed with which deals are reached, as well as the number and the Pareto efficiency of the deals reached.

In the context of bundling, the distinction between complementary and non-complementary goods is important. In the case of complementary goods, the valuation of a bundle is higher than the sum of the valuations of the individual goods. Bundling complementary goods clearly results in higher gains from trade and is therefore mutually beneficial. Firms usually know beforehand which goods do and which do not complement one another (e.g., bicycle and bicycle tier, copier and toner, etc.), and for complementary goods they will make straightforward bundling decisions accordingly. For an important subclass of non-complementary goods---so-called additively separable goods---the bundle valuation is obtained by just adding up the individual valuations. In that case, the way in which bundling may be advantageous is less clear: it depends on the shop's and the customer's valuations. The shop may enjoy economies of scale or scope in the production or distribution of goods, while the customer's valuations for different goods may be correlated (see \cite{Mankila:1999} and the references cited therein). It is this setting of additively separable goods that we will focus on in the current paper. A cable provider with TV, phone, internet, and pay TV services offers an example of additively separable goods. Another example is the common practice of mobile phone operators in Europe to offer prepaid subscriptions for fixed amounts of SMS, long-distance minutes, international calls, and other services. A final example is an online news provider selling various news items in relatively independent categories such as sports, finance, culture, and science.

For numerous real world applications---like the ones mentioned above---the number of individual goods to be bundled, $n$, is relatively small. In this paper we will also consider only small values of $n$ (say $n \leq 10$), for which aggregate knowledge still greatly facilitates the process of finding attractive alternative bundles during a negotiation process. For example, with $n = 10$, there are $2^n-1=1023$ possible bundle configurations, so facilitating the search process among all those bundles is highly valuable. On the other hand, obtaining the desired aggregate knowledge is still manageable, since with additively separable goods this only requires information about customers' valuations for individual goods, and not for all possible bundles.

The next section provides a high-level overview of the interaction model. In Section~$\ref{sec:usermodel}$ we introduce relatively mild conditions on the seller's and his customers' preferences. Based on these conditions, Section~$\ref{sec:model}$ develops a procedure for finding the most promising alternative bundles. In order to test the performance of our system, we used it in interactions with simulated customers. Section~$\ref{sec:simulation}$ presents our computer experiments and discusses the results. Conclusions follow in Section~$\ref{sec:discussion}$.

\section{Overview}\label{sec:overview}

This section gives an overview of the interaction between the shop and the customer, as they try to negotiate an agreement about the price and the composition of a bundle of goods. The shop sells a total of $n$ goods, each of which may be either absent or present in a bundle, so that there are $2^n - 1$ distinct bundle-configurations containing at least $1$ good. In the current paper, we use $n = 10$. A negotiation concerns a bundle (configuration), together with a price for that bundle, and it is conducted in an alternating exchange of offers and counter offers \cite{Rubinstein:1982}, typically initiated by the customer. An example of such a practice may involve the sale of bundles of news items in categories like politics, finance, economy, sports, arts, etc.

We develop a procedure that a seller can use to find mutually beneficial alternative bundles during the negotiation about a given bundle, so that alternative bundles may be recommended whenever the negotiation about the given bundle stalls. Specifically, the procedure finds \emph{Pareto improvements} by changing the bundle content.\footnotemark[\theparetoFootnote] It uses information specific to the current negotiation process as well as aggregate knowledge.\footnote{Aggregate knowledge may be obtained by analyzing past sales data or provided by marketing experts.} The ongoing negotiation is analyzed to determine \emph{when} an alternative bundle is needed, and both the ongoing negotiation process and the aggregate knowledge are used to assess \emph{which} bundle to recommend.

A customer can explicitly reject a suggested bundle by specifying a counter offer with a different bundle content (e.g., the previous one), and she can implicitly reject a suggested bundle by offering a low price for it. In the current paper, only implicit rejection is allowed: customers only specify the bundle content for the opening offer, and thereafter only the seller can change the bundle content of an offer. This is to ease the description of our model and solutions. The possibility for customers to explicitly reject or change the bundle content can be easily incorporated in our model and solutions, however.

\begin{figure}[ht]
\center
\centering \resizebox{\textwidth}{!}{\includegraphics*{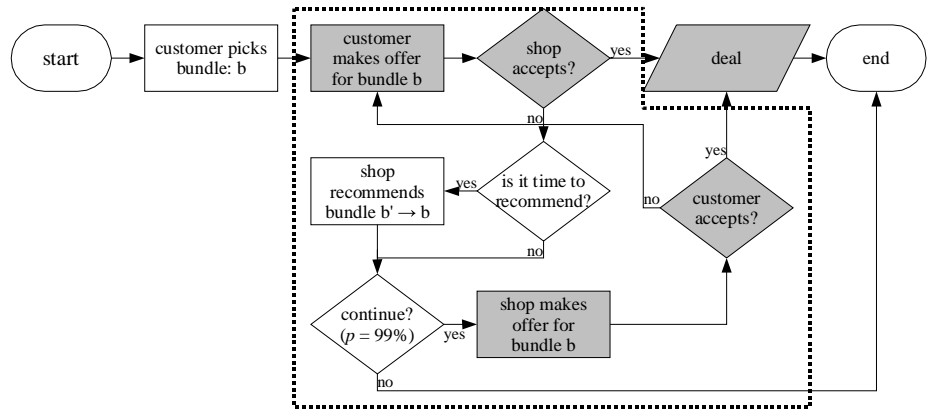}}
\caption{A flowchart describing the integration of recommendation in a shop and a customer's alternating exchange of offers and counter offers.}
\label{fig:flowchart} 
\end{figure}

Figure~$\ref{fig:flowchart}$ provides a high-level overview of the interaction between a shop and a customer. The shaded elements are part of the actual negotiation---the exchange of offers.
The process starts with the customer indicating her interests, by specifying the bundle they will initially negotiate about. After that, they enter into a loop (indicated by the dotted line) which ends only when a deal is made, or with a $1\%$ exogenous probability. (We do not model bargainers' impatience explicitly; therefore we need an exogenous stopping condition, which specifies the chance of bargaining breakdown.) In the loop, the customer makes an offer for the current bundle $b$, indicating the price she wants to pay for it. The shop responds to this offer either by accepting it, or by considering a recommendation. In any case, conditional upon the $99\%$ continuation probability, the shop also makes an offer, either for the current bundle $b$ or for a recommended bundle $b'$ (which then becomes the current bundle). 

In the model, the \emph{valuations} of the customers and the seller are expressed as \emph{monetary values}. The \emph{utilities} of deals are expressed as strictly monotonic one-dimensional transformations of valuations. In the simplest form, this would be the difference between the valuation of the bundle and the negotiated price. The agents are interested in obtaining a deal with optimal utility (``net monetary value''). See Section~$\ref{sec:usermodel}$ for details.

\section{Preference Model}\label{sec:usermodel}

\subsection{Informal Discussion}

The essence of our model of valuations and preferences lies in the assumption that customers and seller order bundles based on their \emph{net monetary value}; the bundle with the highest net monetary value is the most preferred bundle. A customer's net monetary value of a bundle is equal to the customer's valuation of the bundle (expressed in money) minus the bundle price and the seller's net monetary value is equal to the bundle price minus the seller's bundle valuation (also expressed in money).

Given the above assumption and the assumption that a customer wants to buy at most one bundle (within a given time period), Section~$\ref{subsection:formal_discussion}$ shows that any deal involving the bundle with the highest \emph{gains from trade} is Pareto efficient. We can now specify which is the best bundle for the seller to advise: faced with the problem of recommending one bundle out of a collection of bundles, the ``best'' bundle to recommend is the bundle with the highest expected gains from trade; this bundle Pareto dominates all other bundles. (Section~$\ref{subsection:formal_discussion}$ can be skipped upon first reading.)

\subsection{Formal Discussion}\label{subsection:formal_discussion}

Before being able to more formally state the results, some notation is necessary. Let $N\subset \mathbb{N}$, with $n=|N|$, denote the collection of $n$ individual goods and $2^N$ denotes the power set of $N$ (i.e., the collection of all subsets of $n$), then $B=2^N\setminus\{\emptyset\}$ denotes the collection of all possible bundles. Furthermore, let $P = \mathbb{R}$ denote the collection of all possible bundle prices.\footnote{Negative prices may not be realistic, but we want to make as few behavioral assumptions as possible. For the results the possibility of negative prices is not problematic (see footnote $\ref{footnote:pos_prices}$).} The customer and the seller attach the monetary values of $v_c(b)$ and $v_s(b)$, respectively, to a bundle $b\in B$ (with $v_c(b), v_s(b) \in P$). The function $x_j : B \times P \mapsto \mathbb{R}$ with $j \in \{c,s\}$ denotes the \emph{net} monetary value for bundle $b$ and bundle price $p$: $x_c(b, p) = v_c(b) - p$ and $x_s(b,p)= p - v_s(b)$ denote the customer's and the seller's net monetary values, respectively.

We assume that the customer's and the seller's utility for consuming bundle $b$ for a price $p$, denoted by $u_j(b, p)$ with $j\in \{c,s\}$, can be expressed as the composition function $g_j \circ x_j(b, p)$ with $j\in\{c,s\}$ and $g_j:\mathbb{R} \mapsto \mathbb{R}$. For $g_j$ we assume that $\frac{dg_j(x)}{dx} > 0 \mbox{ for all }x\in \mathbb{R}$ and $j \in \{c,s\}$.

Given the customer's and seller's monetary values, we define a useful subset $B^*$ of $B$ as follows: $B^* \equiv \arg\max_{b\in B}( v_c(b) - v_s(b))$, that is, $ B^*$ represents the collection of bundles with the highest gains from trade. We are now ready to introduce the following proposition.

\begin{prop}\label{prop:paretoEffBundle}
A deal $(b, p)$ with $b \in B$ and $p \in P$ is Pareto efficient if and only if $b\in B^*$.
\end{prop}

\begin{rem}
A deal $(b, p)$ is Pareto efficient if there is no $(b', p')$ such that $u_j(b, p) \leq u_j(b',p')$ for all $j\in \{c,s\}$ and the inequality is strict for at least one $j$.
\end{rem}

Proposition~\ref{prop:paretoEffBundle} means that a deal is Pareto efficient if and only if it entails a bundle with the highest gains from trade. For the proof of this proposition the following lemma is very useful.
\begin{lem}\label{lemma:inequality}
For any two deals $(b^*, p^*)$ and $(b, p)$ with $p^*, p \in P$, $b^* \in B^*$, and $b \in B \setminus B^*$ we have $x_c(b, p) < x_c(b^*, p^*)$ or $x_s(b, p) < x_s(b^*, p^*)$.  
\end{lem}

\begin{proof}
We prove the above lemma by contradiction. Suppose that for any $b^* \in B^*$ and $b \in B \setminus B^*$ we have $x_c(b, p) \geq x_c(b^*, p^*)$ and $x_s(b, p) \geq x_s(b^*, p^*)$. A necessary conditions for this to hold is that $v_c(b) - v_s(b) \geq v_c(b^*) - v_s(b^*)$. However, $b^* \in B^*$ and $b \in B \setminus B^*$ means, by definition of $B^*$, that $v_c(b) - v_s(b) < v_c(b^*) - v_s(b^*)$.
\end{proof}

We are now ready to prove proposition~$\ref{prop:paretoEffBundle}$.

\begin{proof}

\begin{enumerate}
\item (If) Pick any $j \in \{c, s\}$. Suppose that $j$'s position improves by moving from any deal $(b, p)$ with $b \in B^*$ to $(b', p')$, that is, $u_j(b, p) < u_j(b', p')$. It then suffices to show that the opponent denoted by $j'$ will always be made worse off, that is, $u_{j'}(b, p) > u_{j'}(b', p')$. From the properties of $g_j$ and $g_{j'}$ it follows that a bargainer's position improves/worsens whenever the net monetary value increases/decreases. Since $j$'s position improves, it follows from lemma~$\ref{lemma:inequality}$ that $j'$ is made worse off whenever $b \in B \setminus B^*$. Moreover, if $b^*, b \in B^*$ then the gains from trade remain unchanged, hence $j'$ is made worse off.
\item (Only if) We will prove this part by contradiction. Suppose that $b \notin B^*$ with the price being any $p \in P$. Pick any $b' \in B^*$ and set the bundle price to $p' = p + v_s(b') - v_s(b)$, so that $p' - v_s(b') = p - v_s(b)$. It follows from $p \in P$ that $p' \in P$ (recall that $P = \mathbb{R}$)%
\footnote{\label{footnote:pos_prices}If we choose to a priori rule out $p < 0$ and $v_j(b) < 0$ (for $j \in \{c, s\}$ and all $b \in B$), then $p \geq v_s(b)$ should hold because otherwise the seller will not be willing to sell the bundle in the first place. Consequently, $p' \in P$ still holds.}
and the properties of $g_s$ that the seller is indifferent between the deals $(b, p)$ and $(b', p')$. Also, it follows from lemma~$\ref{lemma:inequality}$ and the properties of $g_{c}$ that the customer is made better off. That is, any $b' \in B^*$ Pareto dominates $b \notin B^*$. Thus $b \notin B^*$ cannot be a Pareto efficient solution.   
\end{enumerate}

\end{proof}

\section{Recommendation Mechanism}\label{sec:model}

The idea is to develop a mechanism for a seller to find Pareto improvements by changing the bundle content during a negotiation. The mechanism we propose contains two subprocedures. The first procedure---telling the shop when to recommend---monitors the negotiation process and determines when to pass control to the second procedure, which generates recommendations based on aggregate knowledge and the ongoing negotiation process. Figure~$\ref{fig:flowchart}$ already showed the interaction between these two procedures; they are discussed in more detail in Sections~$\ref{subsec:model-when2advise}$ and $\ref{subsec:model-recommending}$, respectively.

\subsection{Deciding \emph{When} to Recommend}\label{subsec:model-when2advise}

The shop needs a procedure for deciding when he should recommend negotiating about a different bundle. The obvious input for this decision is the progress of the current negotiation process, which can be described as a sequence of offers and counteroffers. An offer $O$ contains a bundle definition and a price: $O = (b, p)$ with $b \in B$ and $p \in P$. ($B$ and $P$ denote the collections of all possible bundles and prices, respectively.) Let $h = (O(1), O(2), \ldots, O(k))$ denote a finite history of offers ($k$ is a natural number), where $O(i + 1)$ is the counter offer for $O(i)$, for all $i < k$. Furthermore, let $H$ denote the universe of all possible finite offer sequences (thus $h\in H$). The problem of when to advise can now be specified as the mapping $f: H \mapsto \{\mathrm{yes}, \mathrm{no}\}$, where ``$\mathrm{yes}$'' means: recommend a new bundle.

We construct a heuristic for $f$ based on the assumption that there is a probability of not reaching a deal with a customer (e.g., a break off, endless repetition, or deadline): the longer the negotiation is expected to take, the less likely a deal is expected to become. Furthermore, as a deal becomes less likely, the incentive for the shop to recommend negotiation about an alternative deal should increase. Given the seller's bargaining strategy then, our heuristic extrapolates the time the current negotiation process will need to reach a deal, from the pace with which the customer is currently giving in. More precisely, if we let $O = (b, p)$ and $O' = (b, p')$ denote the customer's current and previous offers for bundle $b$, then $\Delta t$, the predicted remaining number of negotiation rounds needed to reach a deal, is defined as follows:
\begin{equation}
\Delta t= \frac{v_s(b) - p'}{p - p'},
\end{equation} 
where $v_s(b)$ denotes the seller's monetary value for bundle $b$. The higher $\Delta t$, the higher the likelihood of a recommendation. Specifically, the probability of a recommendation depends on $\Delta t$ as follows:
\[
pr_{\mathrm{recommendation}} = 1 - \exp(-0.25 \Delta t),
\]
which means that the probability that the shop recommends an alternative bundle approaches 1 as $\Delta t$ increases.

\subsection{Deciding \emph{What} to Recommend}\label{subsec:model-recommending}

Our mechanism combines aggregate knowledge (obtained from the analysis of sales data and (anonymous) data on previous negotiations, market research, or expert knowledge) with data on the ongoing bargain process, to recommend bundles to customers while negotiating with them. Suppose, for example, that a customer offers to buy a bundle $b$ at a price $p$. When a recommendation is needed (see Section~$\ref{subsec:model-when2advise}$) the idea is to select from within the ``neighborhood'' of bundle $b$, the bundle $b'$ that maximizes $E[v_c(b') - v_s(b') | v_c(b) \geq p]$, the expected gains from trade---given that a customer is willing to pay at least the price $p$ for bundle $b$. Since the seller knows its own monetary value for bundle $b'$, $v_s(b')$, the aim is really to maximize $E[v_c(b') | v_c(b) \geq p] - v_s(b')$. The difficulty here lies in estimating the customer's expected valuation of the bundle:
\begin{equation}
E[v_c(b') | v_c(b) \geq p] = \sum_{i \in P} pr(v_c(b')=i| v_c(b) \geq p),  
\end{equation}
where $pr(v_c(b') = i | v_c(b) \geq p)$ denotes the probability that the customer's valuation for bundle $b$ is equal to $i$, given that she is willing to pay at least $p$ for bundle $b$. (To simplify notation we will write $E[v_c(b') | b]$ instead of $E[v_c(b') | v_c(b) \geq p]$.)

Aggregate knowledge can provide an estimation of $E[v_c(b')|b]$. Given that the shop sells $n$ individual goods, there are $2^n - 1$ possible bundles containing at least $1$ good. To determine $E[v_c(b')|b]$ for all possible bundle pairs, requires---worst case---an order of $(2^n)^2$ estimations. When the customer's valuation for a bundle is just the sum of her valuations of the individual goods comprising the bundle, as it is assumed in the current paper, this complexity is reduced significantly. Given that a customer's valuation of bundle $b$, $v_c(b)$, is simply the sum of the valuations of the goods comprising bundle $b$,
\begin{equation}
E[v_c(b') | b] = \sum_{i \in b'} E[v_c(i) | b].
\end{equation}
This requires at worse ``only" $n \cdot 2^{n}$ estimations of conditional expectations, which is manageable for $n = 10$, as in the current paper. For larger values of $n$ it quickly becomes infeasible to estimate these expectations. In that case, it is necessary to approximate most of the conditional probabilities based on a limited number of explicit estimations. In this paper the focus lies on applications where $n\cdot 2^{n}$ is still a manageable number, so discussing approximation techniques is beyond the scope of this paper.

\subsubsection{Generating Recommendations}\label{subsubsec:recommendations}

A customer initiates the negotiation process by proposing an initial bundle and offering an opening price: let $O(0)=(b, p)$ denote the customer's opening offer (with $b \in B$ and $p \in P$). The shop stores the bundle proposed by the customer as the customer's ``interest bundle,'' in the neighborhood of which he searches for promising alternative bundles to recommend if, at any time, the shop decides he should make a recommendation (see Section~$\ref{subsec:model-when2advise}$). This neighborhood of bundle $b$, $\mathit{Ng}(b)$, is defined as follows.
\begin{equation}\label{b_advice_set}
\mathit{Ng}(b) \equiv \{b' \in B : (b' \subset b\mbox{ and } |b'| + 1 =|b|)\mbox{ or }( b' \supset b\mbox{ and } |b'| - 1 =|b|) \},
\end{equation}
In other words, $\mathit{Ng}(b)$ contains the bundles which, in binary representation, have a Hamming distance to $b$ of $1$.\footnote{Remember that each bundle can be represented as a string containing $n$ bits indicating the presence or absence in the bundle, of each of the shop's $n$ goods.} The advantage of advising bundles within the neighborhood of $b$ is that the advice is less likely to appear haphazard.

Having defined a bundle's neighborhood, let the ordered set $A$ denote the so-called ``recommendation set,'' obtained by ordering the neighborhood $\mathit{Ng}(b)$ on the basis of the estimated expected gains from trade of all the bundles $b'$ in bundle $b$'s neighborhood, $\hat{E}[v_c(b') | b] - v_s(b')$, where $\hat{E}$ denotes the estimation of $E$. Let $\bar{A}$ denote the unordered set of previously proposed bundles.

To recommend a bundle $b_k$ (the $k^\mathit{th}$ recommendation, with $k \geq 1$), our mechanism removes the first bundle from $A$, adds a price to it and proposes it as part of the shop's next offer, and then adds it to $\bar{A}$. Depending on the customer's counter offer for bundle $b_k$, a number of additional bundles may be added to the advice set: if the customer's response is very promising (to be defined below), bundle $b_k$ will be taken as the customer's \emph{new} interest bundle (in the neighborhood of which the search continues), and the bundles in the neighborhood of $b_k$ are added to $A$.

Intuitively, in order to determine how promising a bundle $b_k$ is in terms of its potential for generating gains from trade, the shop needs to compare the net monetary value of the new bundle $b_k$, with the current highest net monetary value among all previous bundles. A relatively large improvement over the currently most promising bundle---the shop's estimation of the customer's `interest bundle'---will cause the shop to update his estimation of the customer's `interest bundle'. However, because the shop does not know the customer's valuation for a bundle, he simply compares offered and asked prices for bundles.

To specify this in more detail, let $O^c_t$ denote the sequence of offers placed by the customer up until time $t$, and let $\mathit{max}(O^c_t)$ specify the customer's past offer with the highest difference between the customer's offered and the shop's asked price. Then the shop will determine the impact of recommending bundle $b_k$ by comparing the customer's current offer for bundle $b_k$, $O(t + 1)$ with that of offer $\mathit{max}(O^c_t)$, from the perspective of his own bid for bundle $b_k$. For this purpose, the shop uses the function $\mathit{sign}: \mathbb{R} \times \mathbb{R} \mapsto \{0, 1, 2\}$. If we let $\mathit{max}(O^c_t) = (b', p_c')$, the customer's current offer $O(t + 1) = (b, p_c)$, and the shop's bids for bundles $b$ and $b'$ be $O(b',p_s')$ and $O(b,p_s)$, then
\begin{equation}
\mathit{sign}_{b, b'}(p, p')=
    \left\{
        \begin{array}{lll}
            2 & \mbox{if } \frac{p_c - p_s}{p_c' - p_s'} > (1 + \mathrm{threshold})\\
            1 & \mbox{if } 1 \leq \frac{p_c - p_s}{p_c' - p_s'} \leq (1 + \mathrm{threshold})\\
            0 & \mbox{otherwise}
        \end{array}
    \right..
\end{equation}

If $\mathit{sign}(p, p') = 2$, then the shop's assessment of the customer's interest bundle is updated to be $b_k$: the customer's response is promising enough to divert the search towards the neighborhood of $b_k$, and add that neighborhood to $A$ such that the first elements of $A$ all lie in the neighborhood of $b_k$. That is, the first element of $A$ becomes the bundle $b' \in \mathit{Ng}(b_k)$ with the maximum difference $\hat{E}[v_c(b') | b_k] - v_s(b')$, the second element of $A$ becomes the bundle $b''$ with the second highest difference $\hat{E}[v_c(b'') | b_k] - v_s(b'')$, and so on. In addition, duplicates are removed from $A$, as are bundles already present in $\bar{A}$. In case $\mathit{sign}(p, p') = 1$, the customer's response is promising enough to continue negotiating about the current bundle $b_k$, but not promising enough to change the assessment of the customer's interest bundle, and if $\mathit{sign}(p, p') = 0$, the proposed bundle was not promising at all and the seller will immediately make the next recommendation.

\section{Numerical Experiments}\label{sec:simulation}

In order to test the performance of our proposed mechanism, we implemented it computationally, and tested it against many simulated customers. Valuations for the shop and the customers were drawn from random distributions. First we describe how we handled the approximation process and how we implemented negotiations in the simulation, and then we present our experimental design and simulation results.

\subsection{Modeling $E[v_c(b')|b]$}

In the experiments we abstract away from actually learning $E[v_c(b')|b]$, for example from sales data. Instead we derive these conditional expectations directly from the way we specified the underlying stochastic process.

As explained earlier, we assume additively separable customer preferences. To compute the customer's valuation for a bundle $b$ we simply add up her valuations for the individual goods that constitute the bundle:
\begin{equation}
v_c(b)=\sum_{i \in b}v_c(i).
\end{equation}
Let $N$ denote the collection of all the individual goods from which bundles are constructed, with $|N|=n$. Suppose we specify the joint probability density function of the customer's valuations for the individual goods, $pr(z_1, \ldots, z_n)$, as an $n$-variate normal distribution. Let the vector $\mathbf{\mu}=(\mu_1, \cdots , \mu_n$) denote the mean of the distribution and let the matrix $\mathbf{\Sigma}=[\sigma_{ij}]$ denote the covariance matrix. Then $pr(z_1, \ldots, z_n) \sim N[\mathbf{\mu}, \mathbf{\Sigma}]$.

The joint probability mass function of all bundle valuations, $pr(z_1, \ldots, z_{2^n})$, is simply a linear transformation of $pr(z_1, \ldots, z_n)$. Since a linear transformation of a multivariate normal distribution is also a multivariate normal distribution \cite{green:1993}, we have
\begin{equation}
pr(z_1, \ldots, z_{2^n}) \sim N[\mathbf{T\mu},\mathbf{T\Sigma T'}],
\end{equation}
where the matrix $\mathbf{T}$ specifies the linear transformation (a row in $\mathbf{T}$ specifies a bundle in binary representation). Given $N[\mathbf{T\mu},\mathbf{T\Sigma T'}]$ we can derive the value of $E[v_c(b')|b]$ for any bundle pair. This approach implies that we simply hand the shop the distributions underlying customers' valuations. 

\subsection{Modeling Negotiations}\label{subsec:simulation-modelingNegotiations}

Besides setting customer preferences it is necessary to specify how the shop and the customer actually negotiate. To allow initiation of the negotiation process by the customer, we assume that the customer starts negotiating about an initial bundle $b_{\mathit{init}}$.  In order to give the shop some room for improvement, we initialize the customer's initial bundle as the bundle containing all the goods for which her valuation is lower than her average valuation across all goods. Although this seems to make it very easy for the shop to make an improvement, bare in mind that performance refers to gains from trade, which depends on both the customer's and the shop's valuations. Besides, we measure performance relative to this starting point in our experiments.

\paragraph{Time-dependent Strategy}

\newcommand{\tft}{\textsc{tft}}
\newcommand{\tdf}{\textsc{tdf}}
\newcommand{\tftmf}{\textsc{tftmf}}

For the customer (shop), the time-dependent bidding strategy is monotonically increasing (decreasing) in both the number of bidding rounds ($t$) and her (his) valuation. In particular, a bidding strategy is characterized by the gap the customer leaves between her initial offer and her valuation, and by the speed with which she closes this gap. The gap is specified as a fraction of the bundle valuation and it decreases over time as $\mathit{gap(t)} = \mathit{gap_{init}} \cdot \exp (-\delta t)$, so over time, she approaches the valuation of the bundle she is currently negotiating about. Note that changes in the gap are time-dependent, but not bundle-dependent! This strategy is therefore called ``time-dependent-fraction'' (\tdf).\footnote{We originally implemented a time-dependent strategy that decreases an \emph{absolute} gap over time, but the results from that strategy were qualitatively similar to the fraction strategy's results, so we describe only the latter as they are more intuitive.} The initial gap, $\mathit{gap_{init}}$, and $\delta$ are drawn randomly from a uniform distribution between $[0, 0.5]$ and $[0.1, 0.4]$, respectively. Almost the same holds for the shop's bidding strategy, \textit{mutatis mutandis}. Since $\delta$ already fluctuates for the customer's strategy we do, however, set $\delta = 0.1$ for the shop, in order to reduce the number of jointly fluctuating parameters somewhat. Summing up, the customer (shop) starts her bidding for a bundle at a randomly chosen point between her valuation and (one and a) half her valuation, and gradually approaches her valuation with her bids. 

\paragraph{Tit-for-Tat Strategy}

The time-dependent strategy described above generates bids irrespective of what the opponent does. As an example of a strategy that responds to the opponent, we implemented a variant of tit-for-tat (\tft) \cite{axelrod:1984}. The initial `move' is already specified by $\mathit{gap_{init}}$ like in the \tdf-strategy. If in subsequent moves the utility level of the opponent offer improves, then a fraction $\delta$ of that amount is conceded by the customer. Note that it is the increment in the utility level perceived by the customer. Furthermore, this perceived utility improvement can also be negative. To make the bidding behavior less chaotic, no negative concessions are made. That is, we used a so-called monotone version called tit-for-tat-monotone-fraction (\tftmf) which can never generate a bid with a worse utility than the previous bid.

\subsection{Results}\label{subsec:results}

\paragraph{A Benchmark}

In order to assess the relative performance of the system we conducted the same series of experiments (see below) with a benchmark procedure, which randomly recommends a bundle from the current bundle's neighborhood. That is, the benchmark does not base the order in which it advises the next bundle on the estimated expected gains from trade like our system does.

In our experiments there are $10$ individual goods. We generate the means of the underlying probability density function $pr(z_1, \ldots, z_{10})$ by randomly sampling numbers between $40$ and $250$ without repetition. To exclude negative numbers we then randomly draw the variance of the $10$ marginal distributions from a uniform distribution such that the probability of a negative valuation for an individual good is negligible, viz.\ at most $0.0003$. In order to ensure sufficient differences in valuations between customers, however, we fix the correlation matrix (but not the covariance matrix). To test the robustness of our procedure to quantitative changes in the underlying distributions we conducted a series of experiments with $100$ different distributions $pr(z_1, \ldots, z_{10})$. For each distribution we tested the influence on the system's performance of changes in the ease with which the shop updates his estimation of the customer's ``interest'' (see the discussion of the $\mathit{sign}$ function in Section~$\ref{subsubsec:recommendations}$). The ``threshold'' used in the $sign$ function captures this sensitivity; we experimented with $11$ values between $0$ and $0.5$, with stepsize $0.05$. For each of these settings we simulated negotiations between the shop, with randomly drawn valuations, which were kept constant across negotiations with $100$ customers, each with her valuations drawn randomly from the particular distribution used. The values in the graphs are averages across $100$ customers per distribution, and across $100$ different distributions.

The shop's bundle valuations are not additively separable, due to the follwoing nonlinear pricing strategy. Bundles with a higher than average expected customer valuation---compared to bundle containing the same number of individual goods---are relatively expensive. That is, for expensive bundle it is less likely that customers are actually willing to be the offered good. Similarly, bundles with a lower than expected valuation are relatively inexpensive (compared to bundles of the same size). 

Concerning the legends throughout, the graph for `rounds' gives the number of rounds it took, if a deal was reached, to reach that deal, and the graph for `deals' shows the total number of deals reached in negotiations with the $100$ customers. The `diff' graphs show the difference between our system's and the benchmark's performance. For an explanation of `perc' and `relP,' consider that given the shop's and the customer's valuations and all the possible bundles, there are maximum and minimum gains from trade attainable. The graph for `perc' shows the gains from trade of the final bundle, as a percentage of the difference between the maximum and minimum gains from trade, and `relP' shows this percentage, relative to the difference between the maximum gains from trade and the gains from trade of the initial bundle (the starting point of the negotiation).  As a final remark, bear in mind that the lines in the graphs are included to allow easy identification of patterns, but only the symbols (squares, diamonds, etc.)\ represent actual values measured in the simulation.

\begin{figure*}[ht]
\resizebox{\textwidth}{!}{\includegraphics*[angle=270]{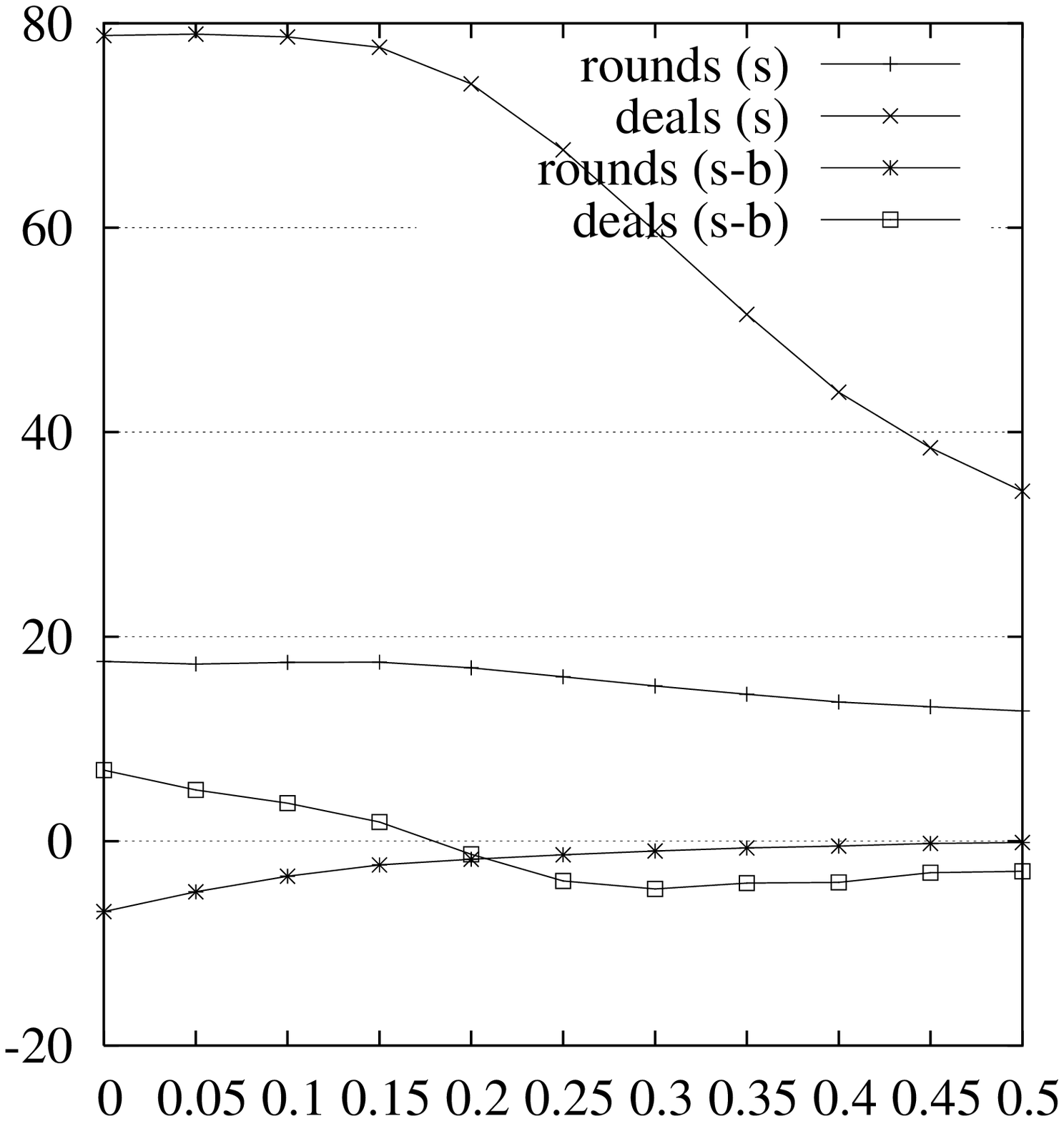} \includegraphics*[angle=270]{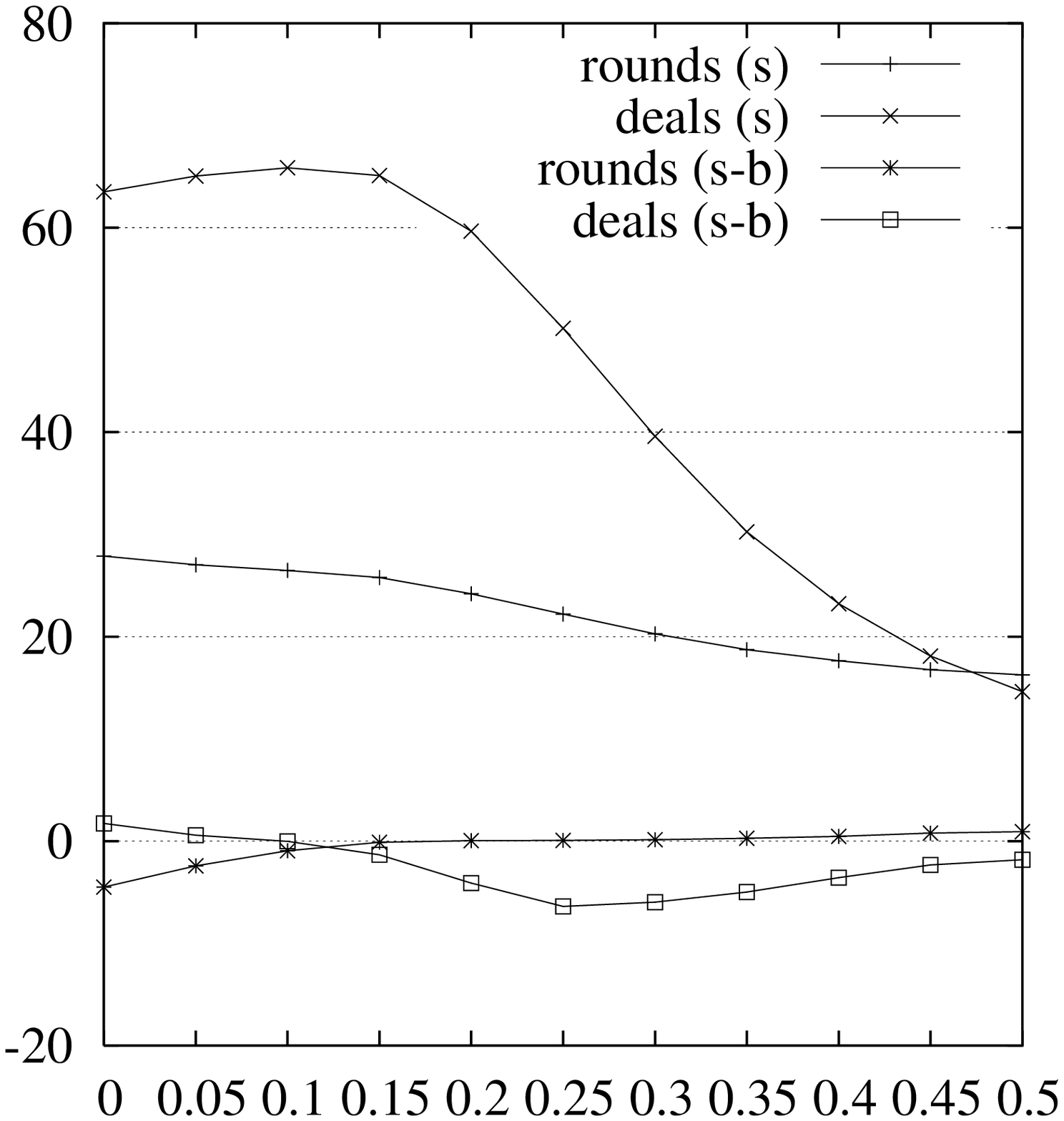} \includegraphics*[angle=270]{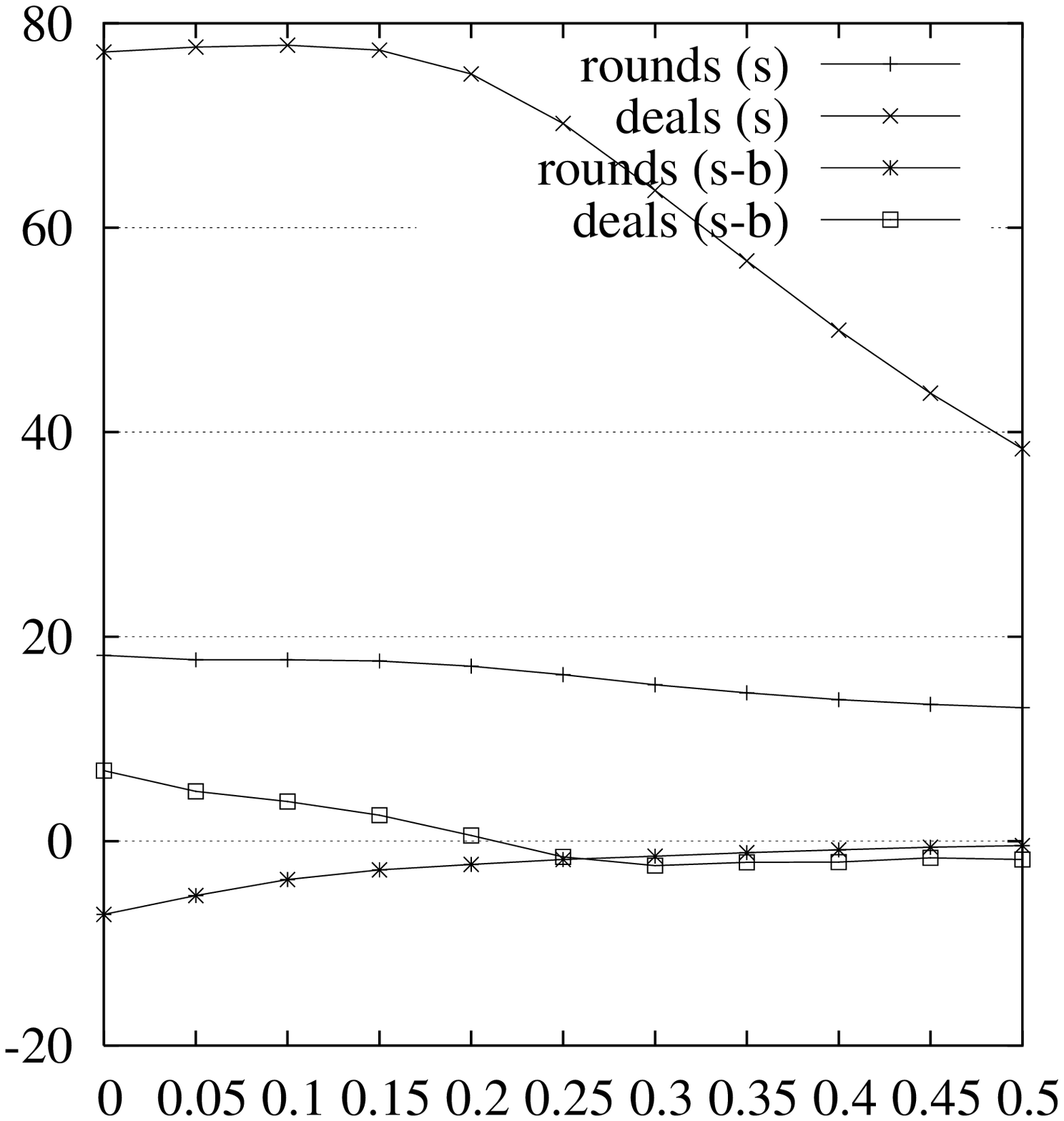}}
\resizebox{\textwidth}{!}{\includegraphics*[angle=270]{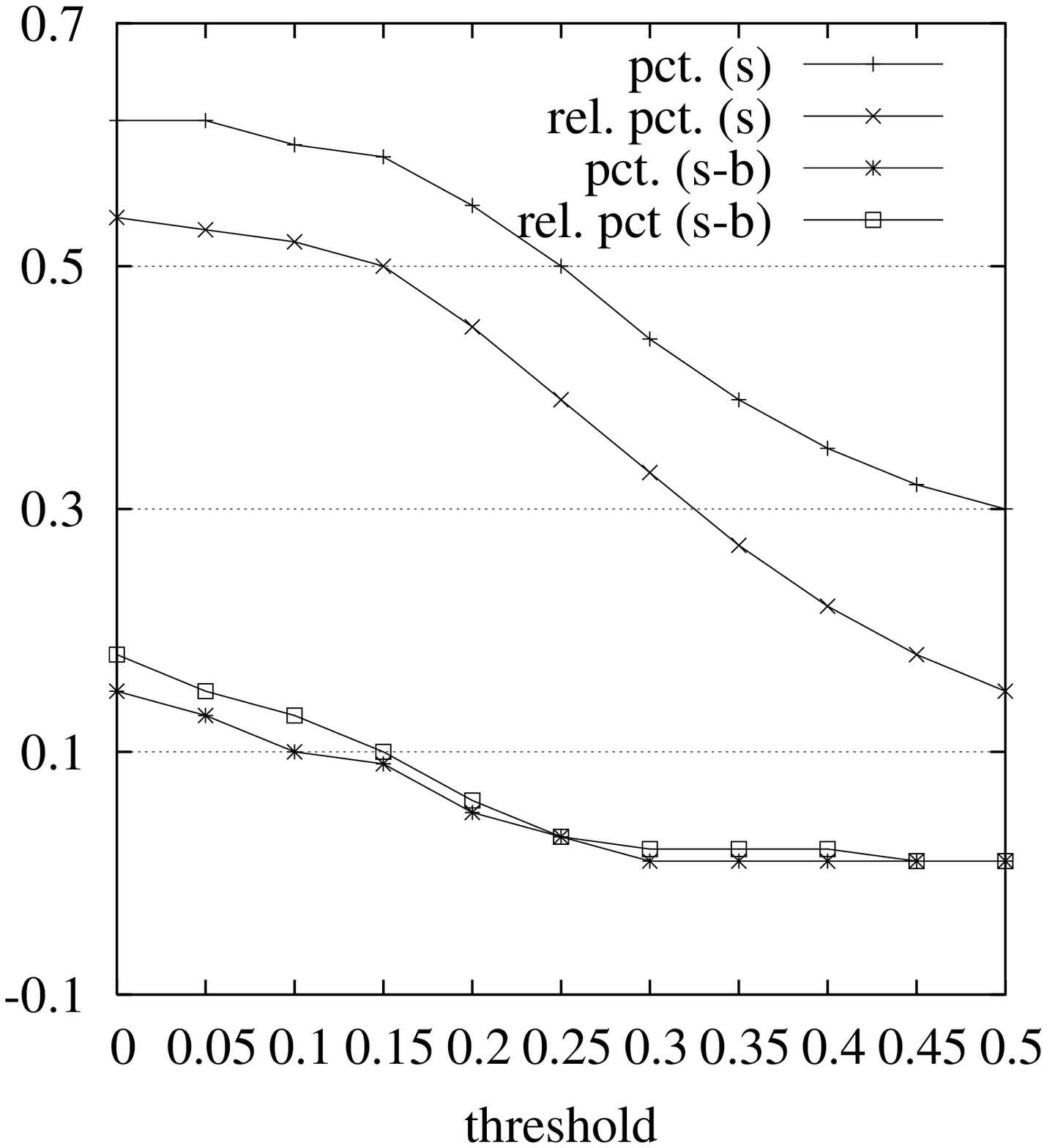} \includegraphics*[angle=270]{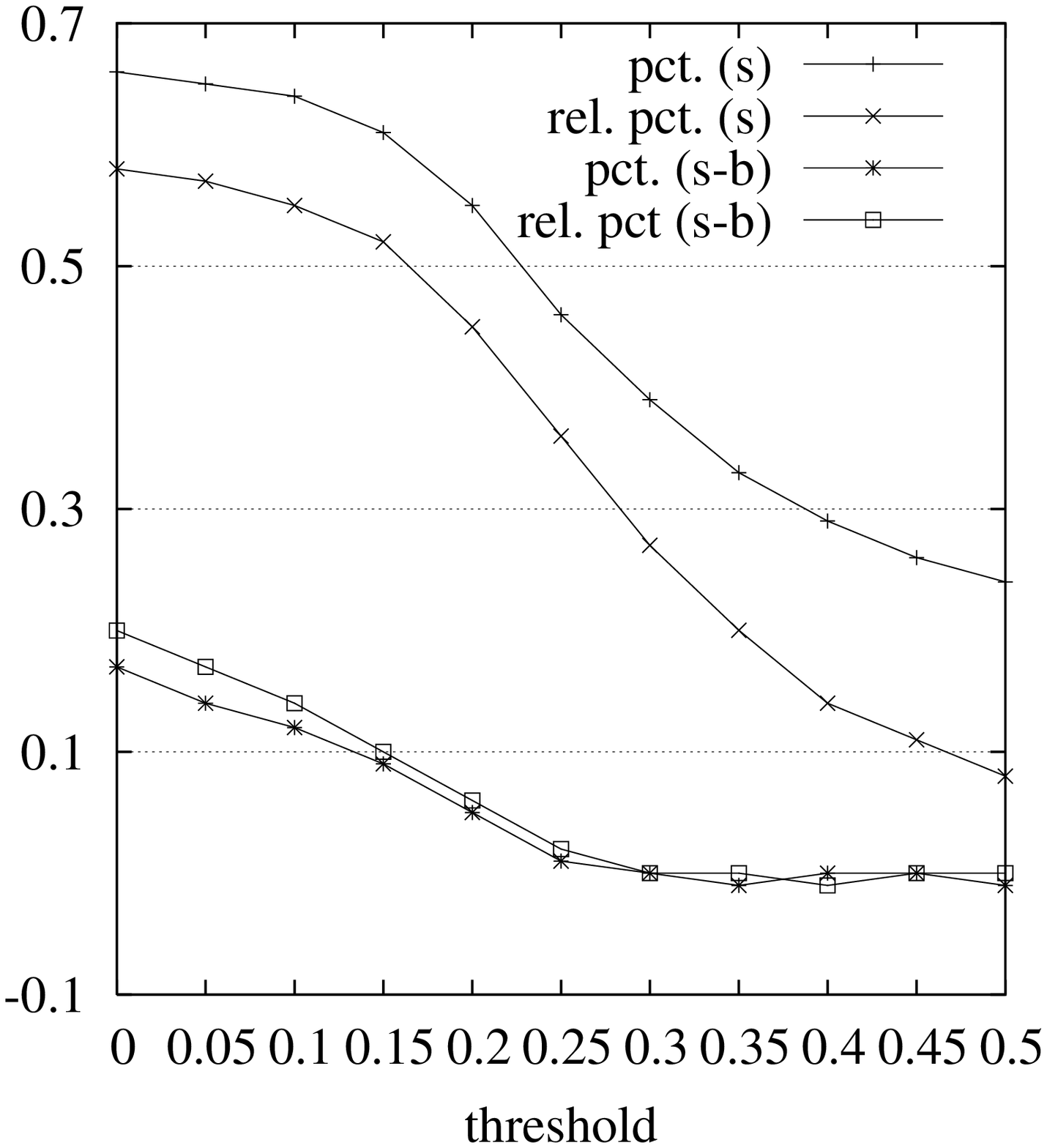} \includegraphics*[angle=270]{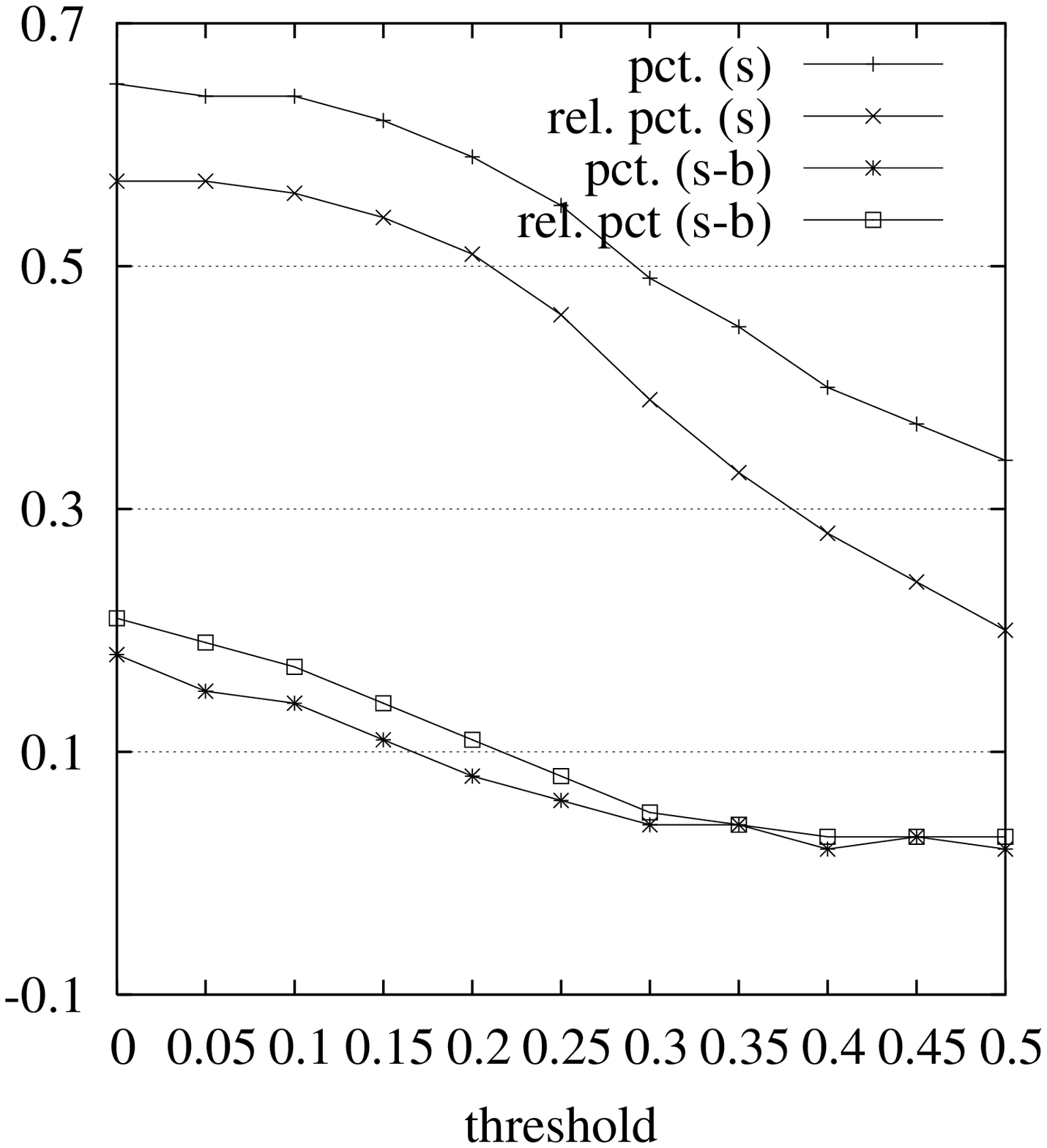}}
\caption{Results for our system (`s'), when the shop uses the \tdf\ strategy with $\delta = 0.1$, and the customers use either the \tdf\ strategy with random $\delta$ (on the left), or the \tftmf\ strategy with random $\delta$ (in the middle) or with $\delta = 1$ (on the right), as described in section~$\ref{subsec:simulation-modelingNegotiations}$. The difference with the benchmark is indicated by the `diff'-graphs.}
\label{fig:td}
\end{figure*}

Figure~$\ref{fig:td}$ reports the results of three series of experiments (see the caption) where we vary the bargaining strategy of the customers. For low thresholds, our system generates roughly $70\%$ of the maximum gains from trade and roughly $60\%$ of the gains of trade attainable given the initial bundle. Its performance in both cases is roughly $20\%$ better than the benchmark. Additionally, more deals are reached and it requires less time to reach these deals than the benchmark.

Initially, larger threshold values improve the performance of the benchmark. As it searches the current interest bundle's neighborhood in a random order, increasing the threshold makes it change the interest bundle less haphazardly, forcing it to focus longer on a particular interest bundle's neighborhood, which may contain better alternatives than the ones it encounters initially. Usually, these better alternatives do exist, and they turn up in such a slower, and more prudent search, leading to more deals as the threshold increases. As the threshold increases further, however, the interest bundle is not updated as easily anymore, thereby blocking the search in regions in the neighborhood of promising bundles. This effect is also visible in the results of our system, which zeroes in on the most promising bundle more quickly. For larger threshold values this effect becomes increasingly more important. Consequently, the difference in performance reduces for larger values; ultimately, for a threshold of $0.5$, there is virtually no distinction in performance.

\section{Conclusions and Future Work}\label{sec:discussion}

We consider the problem of negotiating over both bundle contents and price, which permits a high degree of personalization of bundles to the preferences of customers. We develop a procedure for a seller to search for Pareto improvements in bundle contents, while negotiating about the price of bundles. Computer experiments show how this procedure increases both the speed with which agreements are reached, as well as the number and quality of agreements reached.

In the current paper, we have only considered additively separable consumer preferences. The most important issue currently under investigation is the extension of our procedure to cases involving non-linear preferences.

Another issue concerns the distribution of valuations we used. Such a distribution generates preferences for customers, which may make them buy the shop's (bundles of) goods. Data about such sales, in turn, enable the shop to estimate the distribution underlying his customers' preferences. In the current paper we provided the shop with the aggregate knowledge required for our procedure directly. Even without modeling this process explicitely, we are interested in providing the shop with only an estimate of the real distribution and in testing the robustness of our procedure to variations in the accuracy of the estimate.

\bibliographystyle{splncs}
\bibliography{names,ec}

\end{document}